\documentclass{ws-procs9x6}

\begin{document}

\title{Distribution of Neutron Resonance Widths}

\author{Hans A. Weidenm{\"u}ller}

\address{Max-Planck-Institut f{\"u}r Kernphysik,\\
69120 Heidelberg Germany\\
$^*$E-mail: Hans.Weidenmueller@mpi-hd.mpg.de}

\begin{abstract}
Recent data on neutron resonance widths indicate disagreement with the
Porter--Thomas distribution (PTD). I discuss the theoretical arguments
for the PTD, possible theoretical modifications, and I summarize the
experimental evidence.
\end{abstract}


\bodymatter

\section{Why a Statistical Theory? A Reminder}
\label{aba:sec1}
Isolated resonances observed in the scattering of very slow neutrons
(with energies typically below 10 keV) on medium--weight and heavy
nuclei have a mean spacing of about 10 eV and an average total width
of about 1 eV. The resonances correspond to particle--instable excited
states of the compound nucleus occurring at about 6 to 8 MeV
excitation energy, with several million excited states lying between
the ground state and the resonances. Although the individual quantum
numbers of the resonances may be known experimentally, the number of
states involved is so huge that theory cannot be expected either now
or in the forseeable future to be able to determine the positions,
quantum numbers, and widths of these resonances. The only viable
theoretical access to their properties is a statistical one where the
description of individual resonance properties is replaced by
statements about the distribution of resonance parameters (spacings of
resonance energies, neutron widths, gamma widths). That approach goes
back to E. Wigner and is based upon random--matrix theory, see the
review~\cite{Wei09}.

\section{Purpose}

Random--matrix theory (RMT) predicts that the reduced neutron widths
$\Gamma^{0}_n$ (obtained from the measured partial neutron widths by
removing the appropriate kinematical factor) follow the Porter--Thomas
distribution (PTD) $P(y)$, a $\chi^2$ distribution with one degree of
freedom ($\nu = 1$) given by
\begin{equation}
P(y) = \frac{1}{(2 \pi y)^{1/2}} \exp \{ - y / 2 \} 
\label{1}
\end{equation}
where $y$ is the ratio of $\Gamma^{0}_n$ and its average value. The
PTD was successfully tested~\cite{Boh83} many years ago for the
Nuclear Data ensemble~\cite{Haq82}. However, a recent
test~\cite{Koe10} of the PTD in the Pt isotopes has shown that the PTD
must be rejected there with 99.997 per cent significance. And a
re--analysis~\cite{Koe11} of the Nuclear Data Ensemble (1245 partial
neutron widths) rejects the PTD with 98.17 per cent significance. The
actual analysis and the way these conclusions are arrived at are
discussed in the next talk by P. E. Koehler. Here I focus on a review
of the underlying theory and of the available computational and
experimental evidence, with emphasis on the question: Where does the
theory or the analysis possibly fail? I will not only discuss neutron
widths but summarize (i) the theoretical basis for the random--matrix
prediction and supporting evidence from the nuclear shell model; (ii)
the special assumptions that go beyond RMT and are used to derive the
PTD; (iii) the experimental evidence.

\section{Random--Matrix Theory and Evidence from the Nuclear Shell
Model}
\label{RMT}

RMT is a statistical theory that yields the generic spectral
properties of {\it bound} quantum systems with fixed symmetry. For a
time--reversal invariant system, the Hamiltonian matrix $H$ (dimension
$N \to \infty$) for states with fixed quantum numbers (spin and
parity) is real and symmetric. The matrix elements are assumed to be
random variables. Except for a normalization factor, their
distribution is given by
\begin{equation}
\exp \{ - (N / \lambda^2) \ {\rm Trace} (H^2) \} \ {\rm d} [H] \ .
\label{2}
\end{equation}
Here $\lambda$ is a parameter that has the dimension energy and
determines the mean level spacing. The symbol ${\rm d} [H]$ stands for
the product of the differentials of the independent matrix elements.
Obviously each matrix element has a Gaussian distribution centered at
zero. The distribution~(\ref{2}) is invariant under orthogonal
transformations of Hilbert space and referred to as the Gaussian
Orthogonal Ensemble (GOE). Transforming Eq.~(\ref{2}) to the diagonal
representation of $H$ with eigenvalues $E_\mu$, $\mu = 1,
\ldots, N$ one finds
\begin{equation}
\exp \{ - (N / \lambda^2) \ \sum_\mu E^2_\mu \} \ \prod_{\mu < \nu}
|E_\mu - E_\nu| \prod_\sigma {\rm d} E_\sigma \ {\rm d} [O] \ .
\label{3}
\end{equation}
The factor ${\rm d} [O]$ is the Haar measure of the orthogonal group
in $N$ dimensions. The probability density~(\ref{3}) is the product of
a term which depends only upon the eigenvalues and another that
depends only upon the eigenfunctions. Therefore, eigenfunctions and
eigenvalues are statistically independent. The factor ${\rm d} [O]$
implies that for $N \to \infty$, the projections of the eigenfunctions
onto an arbitrary fixed vector in Hilbert space have a Gaussian
distribution and that, therefore, the squares of these projections
follow the PTD.  Thus, the PTD is based upon orthogonal invariance
only and is seen to be tantamount to a complete mixing of the states
in Hilbert space.  Correlations between eigenvalues are determined by
the factors $\prod_{\mu < \nu} |E_\mu - E_\nu| \prod_\sigma {\rm d}
E_\sigma$ and are not discussed here.

Does the RMT prediction for strong mixing of the eigenstates carry
over to a realistic nuclear model like the shell model? Like RMT that
model is a theory of bound states but differs from RMT in three
important respects. (i) The model is governed by a mean field; (ii)
the residual interaction of the model is a few--body and, usually, a
two--body interaction; (iii) the complexity of angular--momentum
coupling restricts applications of the model to numerical
investigations. As for item (i), the single--particle states within a
major shell are not degenerate. Even in the absence of any residual
interaction, the many--body states of the model are, therefore, not
degenerate either. For instance, in the middle of $sd$--shell, the
resulting many--body spectrum extends over several $10$ MeV. Strong
mixing of all these states is possible only for a sufficiently strong
residual interaction. As for item (ii), in RMT all states interact via
independent matrix elements while a residual two--body interaction has
vanishing matrix elements between all Slater determinants which differ
in the occupation numbers of more that two single--particle states. As
for item (iii), an analytic extrapolation of shell--model results to
very large matrix dimension is not possible. One resorts to schematic
models instead.

\begin{figure}
\begin{center}
\psfig{file=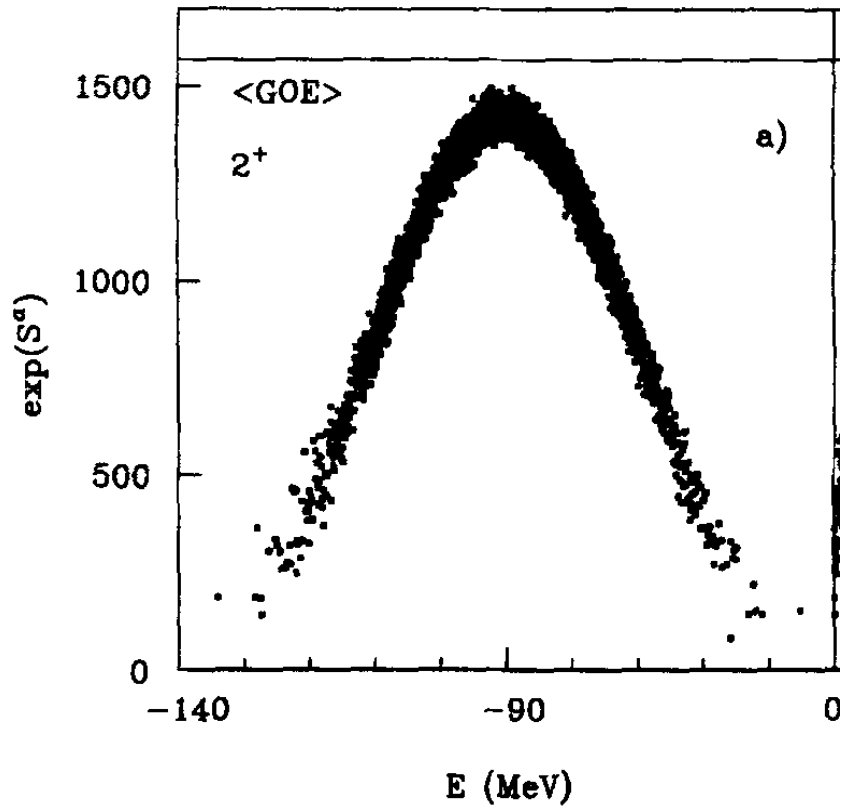,width=4.5in}
\end{center}
\caption{Exponential of the information entropy versus eigenenergy.
The solid horizontal line represents the GOE limit. Mixing of the
eigenfunctions is seen to be nearly complete near the center of the
shell but not in the tails. From Ref.~\cite{Zel96}.}
\label{aba:fig1}
\end{figure}

Early work on the distribution of eigenvector components in the shell
model~\cite{Whi78,Ver79} showed deviations from the Gaussian form that
could be traced to items (i) and (ii) above. There is an overabundance
of small and of very large and a corresponding lack of medium--sized
components reflecting incomplete mixing of the basis states of the
shell model. An extensive test~\cite{Zel96} of statistical properties
of eigenfunctions of the shell model focused on the information
entropy $S_k$, a measure for the mixing of basis shell--model states
of fixed spin and parity. With $W_{k \mu}$ denoting the square of the
expansion coefficient of the normalized $k^{\rm th}$ eigenfunction of
the shell model (ordered by eigenenergy) in the basis $| \mu \rangle$
of shell--model states, $S_k$ is defined as $S_k = - \sum_\mu W_{k
\mu} \ln W_{k \mu}$. For 12 nucleons in the $sd$--shell coupled to
spin $J = 2$ and isospin $T = 0$ ($N = 3276$ states) and interacting
via a standard residual two--body interaction, $S_k$ approaches the
GOE value only in the center of the spectrum, see Fig.~\ref{aba:fig1}.

It is difficult to extrapolate these results to the large matrix
dimension ($N \approx 10^6$) appropriate for resonances at neutron
threshold. French~\cite{Bro81} used the central--limit theorem to
argue that eigenvector components have a Gaussian distribution.
Semi--analytical work~\cite{Pap07} suggests that strong mixing of
eigenfunctions is a generic property of the shell model. For totally
degenerate single--particle states, a generic two--body interaction
leads to complete mixing in the large $N$ limit~\cite{Pap10}. The
critical strength at which a two--body interaction produces chaos in a
system with non--degenerate single--particle states was
estimated~\cite{Jac97} in a schematic model and shown to be given by
the average distance between directly coupled many--body states. For
the $sd$--shell nuclei typical non--diagonal elements of the residual
interaction are of the order of several $100$ keV, i.e, are comparable
to the spacing of single--particle states.

In summary, the evidence from the shell model is somewhat mixed but
points to complete mixing of eigenfunctions in the large $N$ limit.
That is also supported by the fact that in dynamical systems with few
degrees of freedom, RMT is tantamount to quantum chaos, and quantum
chaos (as opposed to integrability) presents the generic case. That
should also apply to nuclei.

\section{Widths}
\label{wid}

RMT is a theory for bound states. Additional assumptions are invoked
when the coupling to the $s$--wave neutron channel is considered. The
eigenstates $| k \rangle$ of RMT become resonances with widths
\begin{equation}
\Gamma_k \propto | \langle \chi(E) | V | k \rangle |^2 \ .
\label{4}
\end{equation}
Here $\chi(E)$ is the antisymmetrized product of the wave function of
the target nucleus and a neutron $s$--wave scattering state at energy
$E$, and $V$ is the nuclear interaction. At face value, $\Gamma_k$ is
the square of the projection of the state $| k \rangle$ onto some
fixed vector and should, therefore, follow the PTD. However, that
argument has the following limitations. (i) The expression~(\ref{4})
is perturbative and holds only for strongly isolated resonances
(average total width $\ll$ average resonance spacing $d$). (ii)
Resonances occur at different neutron energies $E$. The argument
breaks down if the neutron scattering wave function changes
significantly with $E$ over an interval of order $d$. The kinematic
$s$--wave penetration factor $\sqrt{E}$ is a trivial example. (iii)
Within any dynamical model, the wave function of the target nucleus
with $A - 1$ nucleons is an eigenfunction of the same many--body
Hamiltonian that has eigenstates $| k \rangle$ in the $A$--body
system. Specifically, for a shell--model with a random two--body
interaction, both $| \chi \rangle$ and $| k \rangle$ change with the
interaction and are, in a statistical sense, therefore correlated.
That fact violates the assumption that $| k \rangle$ is projected onto
a fixed vector. The correlation comes on top of possible deviations
from the GOE due to the distribution of components of shell--model
eigenvectors as discussed in Section~\ref{RMT}. All three points have
been treated in the literature.

\begin{figure}
\begin{center}
\psfig{file=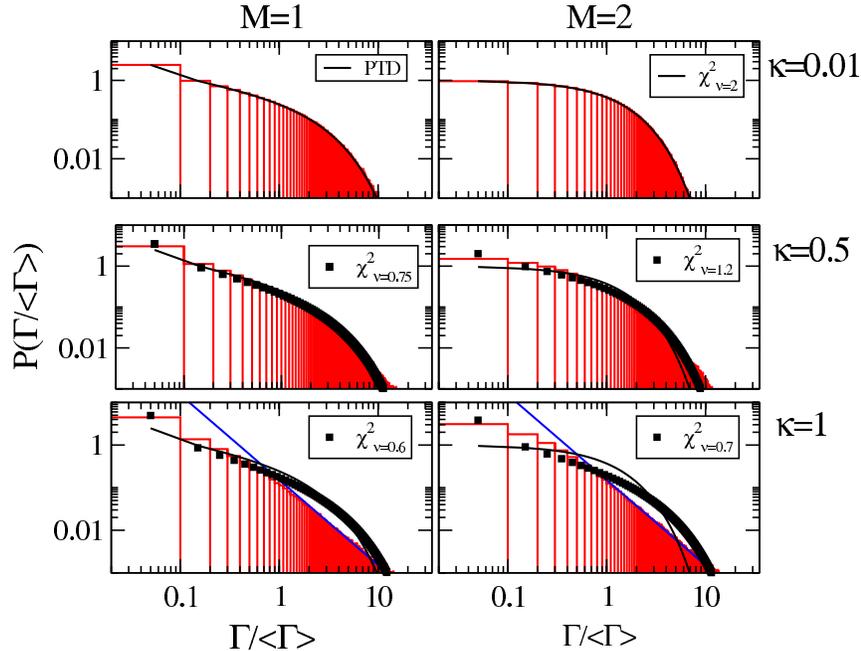,width=4.5in}
\end{center}
\caption{(Color online) Width distributions for $M = 1$ and $M = 2$
channels and for various strengths $\kappa$ of the channel coupling
without removal of the superradiant state. The solid lines give the
PTD. From Ref.~\cite{Cel11}.}
\label{aba:fig2}
\end{figure}

(i) With increasing strength of the coupling to the channel(s), the
PTD is deformed. When an RMT Hamiltonian is coupled to $M$
channels~\cite{Cel11}, deviations from the PTD are already seen for $M
= 1$ and strong coupling. These get even bigger for $M = 2$, see
Fig.~\ref{aba:fig2}. When the superradiant state is removed from the
data, the distribution of the remaining components for the
single--channel case is quite close to the PTD except for intermittent
values of the coupling strength~\cite{Vol11}. For the actual data on
neutron widths analyzed in Refs.~\cite{Koe10,Koe11} the effect seems
marginal.

(ii) A secular dependence of $\Gamma_k$ on resonance energy beyond the
$s$--wave penetration factor $\sqrt{E}$ may be due to a
single--particle resonance near threshold~\cite{Wei10}. As mass number
$A$ increases, single--particle $s$--wave states are pulled into the
shell--model potential. When one of these is close to threshold it
modifies the energy dependence of $\Gamma_k$, and when it is within
$5$ keV from threshold the entire $R$--matrix analysis of the data is
called into question. The Pt isotopes investigated in
Ref.~\cite{Koe10} are close to a maximum of the $s$--wave strength
function. That maximum corresponds to the $4s$ single--particle state
becoming bound. However, the resulting correction factor does not
remove the discrepancy with the PTD distribution for the Pt
isotopes~\cite{Koe11a}, and it is considered unlikely that the $4s$
single--particle state is within $5$ keV from threshold~\cite{Koe11a}.
For the Nuclear Data Ensemble (where most nuclei are far from maxima
of the $s$--wave strength function) the correction factor is
irrelevant. A doorway state above but close to neutron threshold would
likewise cause a secular variation of the $\Gamma_k$ and would distort
the PTD. That possibility seems not to have been studied yet. In any
case, a secular energy dependence of $\Gamma_k$ beyond the $\sqrt{E}$
factor is not expected to occur uniformly for all mass numbers.

\begin{figure}
\psfig{file=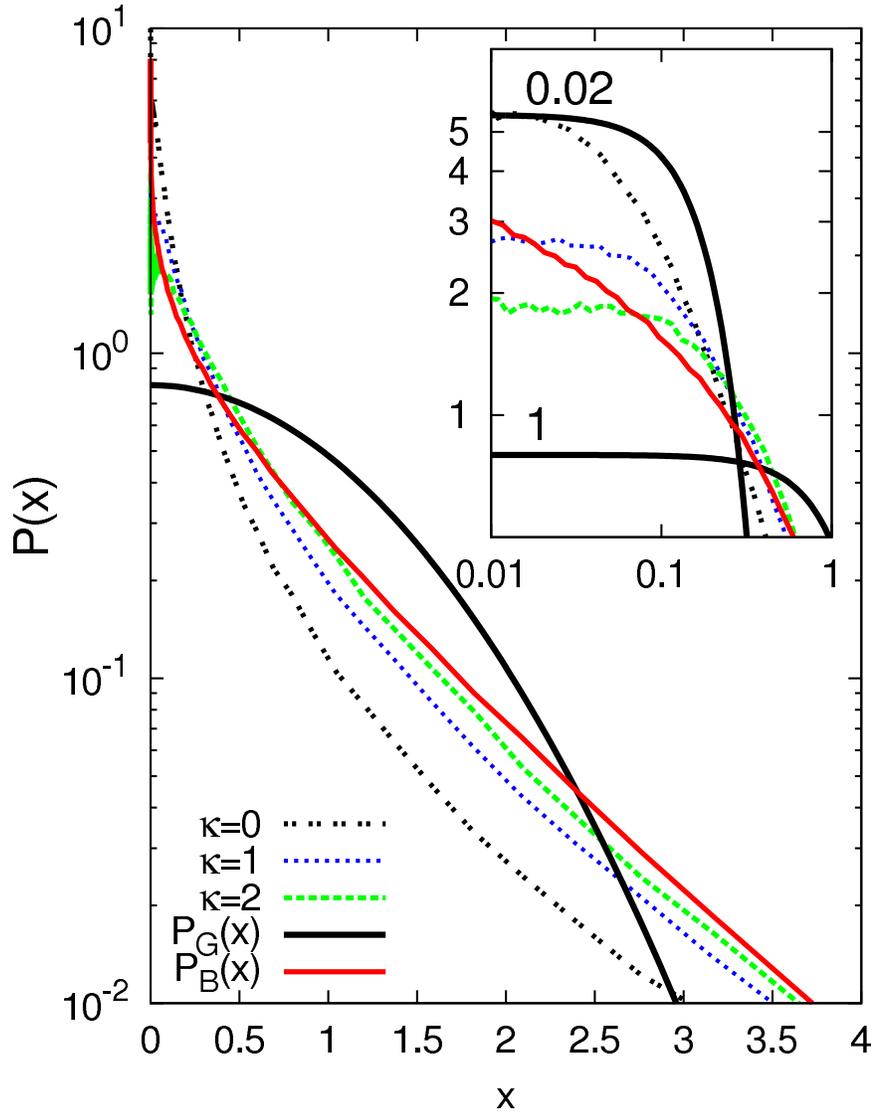,width=4.5in}
\caption{(Color online) Normalized distribution of decay amplitudes
versus magnitude for different coupling strengths $\kappa$. The
solid black line is the Gaussian. From Ref.~\cite{Vol11}.}
\label{aba:fig3}
\end{figure}

(iii) The influence of a correlation between the wave function of the
target nucleus and the resonance eigenstates was studied~\cite{Vol11}
in a schematic model with $7$ particles ($N = 11440$) with a random
two--body interaction and for several strengths of the coupling to the
channel. The amplitude distribution in Fig.~\ref{aba:fig3} (after
removal of the superradiant state) shows deviations from the Gaussian
form for all coupling strengths. For zero coupling (dashed line) there
is no correlation and the effect is the same as discussed for the
shell model in Section~\ref{RMT}. As the coupling is turned on, the
fraction of big components is increased at the expense of the small
ones. The differences with the PTD persist. It is not clear whether
these results persist or are washed out as the matrix dimension is
increased from $10^4$ to $10^6$ or $10^7$ as would be appropriate for
neutron resonances.

\section{Empirical Evidence}

Three types of tests of the distribution of eigenvectors have been
performed.

(i) Statistical independence of eigenvalues and eigenvectors. This
prediction of RMT is not affected by assumptions on the coupling of
the eigenvectors to the channels. A test of that prediction is,
therefore, stronger than is comparison with the PTD. For the Nuclear
Data Ensemble, the value of the correlation coefficient was found to
be consistent with zero~\cite{Boh83}. A more general later
test~\cite{Lom94} supports statistical independence. Both tests
confirm RMT.

(ii) Gaussian distribution of decay amplitudes. The variable in the
PTD is $y_k = \Gamma^0_k / \langle \Gamma^0_k \rangle$. The average
value $\langle \Gamma^0_k \rangle$ may be affected by a few large
values that are statistically irrelevant. A measure that is less
sensitive to $\langle \Gamma^0_k \rangle$ compares the correlation
coefficient of amplitudes (including relative phases) for decay into
two different channels with the correlation coefficient of the
corresponding intensities~\cite{Har84}. For a Gaussian distribution,
the second coefficient is equal to the square of the first. That test
is not available for neutron decay widths (where only the intensities
are known) but has been applied~\cite{Shr87,Shr89} to inelastic proton
resonance data.  Measurements of the (p, p' $\gamma$) angular
correlation yielded 1117 reduced width amplitudes for nuclei in the
mass $50$ region. Within statistics, the square of the amplitude
correlation coefficient was found to be equal to the intensity
correlation coefficient. This is considered a very precise test of the
Gaussian distribution of amplitudes and, thus, of the PTD for widths.

(iii) Neutron widths. There are several problems in the data
analysis. In a given nucleus, the number of neutron resonances is
typically small, so that data from several nuclei must be combined.
Resonances with small widths may remain undetected. Small fluctuations
of the cross section may be mistakenly identified as resonances. Weak
$s$--wave resonances are indistinguishable from strong $p$--wave
resonances. These difficulties are addressed in the next talk by
P. Koehler.

The early analysis~\cite{Boh83} of the Nuclear Data Ensemble showed
very good agreement with the PTD, including the search for the best
$\chi^2$ distribution. Exceptions were found~\cite{For71} in
$^{232}$Th. To eliminate the above--mentioned uncertainties, a
modified analysis~\cite{Cam94} used only reduced widths larger than a
prescribed value. A maximum--likelihood approach was used to determine
the parameter $\nu$ in the distribution $y^{(\nu/2) - 1} \exp \{ - y /
2 \}$ with $y$ as defined above. Here $\nu = 1$ corresponds to the
PTD. For $9$ nuclei with masses $A > 150$ and more than $45$
resonances in each, the error--weighted average value of $\nu$ over
these nuclei is~\cite{Cam94} $\langle \nu \rangle = 0.98 \pm 0.10$, in
agreement with the PTD. The same method but with an energy--dependent
cutoff and $\langle \Gamma^0_k \rangle$ as additional fit parameter
was applied to the $450$ resonances in the Pt isotopes and gave $\nu$
values around $0.6$ so that the PTD was rejected at a significance
level of 99.997 per cent~\cite{Koe10}. For nuclei in the Nuclear Data
Ensemble, the same analysis gave $\nu = 1.217 \pm 0.092$ so that the
PTD is rejected here with a significance of 98.17 per
cent~\cite{Koe11}. It is surprising that the deviations from $\nu = 1$
have opposite signs in the Pt isotopes and in the Nuclear Data
Ensemble, indicating a lack (preponderance) of large widths,
respectively.

\section{Summary and Conclusions}

RMT, general arguments involving the central--limit theorem, entropy
arguments within the shell model, and the large $N$ limit of similar
other models all imply or suggest for matrices of sufficiently large
dimension the validity of the PTD for the distribution of squares of
eigenvector components.

Among possible causes for deviations, strong coupling to channel(s) is
probably not significant for the actual data. A secular energy
dependence of the reduced widths due to threshold effects or to a
doorway state is possibly significant in some (but not all) nuclei.
Insufficient mixing of shell--model states combined with correlations
between target and compound--nucleus eigenfunctions causes deviationd
from the PTD. It is not clear, however, whether these effects would
persist for matrices of realistically large dimension.

Among the various tests, the test for independence of eigenvalues and
eigenvectors supports RMT. The test for the Gaussian distribution of
proton decay amplitudes strongly supports the PTD. In contrast to
earlier work, the distribution of neutron widths from the Pt isotopes
strongly contradicts the PTD and so does the recent re--analysis of
the distribution of neutron widths from the Nuclear Data Ensemble.

The situation is confusing because we have pieces of contradictory
evidence. New data will be helpful and most welcome. But equally
important is an analysis of all existing data that would employ
identical criteria and use all statistical measures available. Such an
analysis should be devoted to the statistical properties of both,
resonance energies and resonance amplitudes (or widths). Since the
Nuclear Data Ensemble was first investigated, new statistical measures
have been introduced, many pieces of data have been added, and
numerous quantum number assignments have been made. All of this should
be taken into account. The situation calls for a concerted effort at
establishing the statistical properties of nuclei.

\end{document}